\documentclass[aps,pra,twocolumn,nopacs,floatfix]{revtex4-2}
\usepackage{epsfig}
\usepackage{graphicx}
\usepackage{dcolumn}
\usepackage{amsthm,amsmath}
\usepackage{comment}
\usepackage[colorlinks]{hyperref}
\usepackage{xcolor}
\usepackage[T1]{fontenc}
\usepackage{multirow}
\usepackage{threeparttable}
\usepackage{mathrsfs}
\usepackage{longtable}
\usepackage{orcidlink}
\usepackage{amsmath}
\usepackage[normalem]{ulem}

\usepackage{resizegather}
\UseRawInputEncoding

\begin{document}

\title{Precise Determination of Electric Dipole Polarizabilities of Sr Clock States Using Linear-Response and Equation-of-Motion Relativistic Coupled-Cluster Methods}

\author{Somesh Chamoli}

\author{Sudipta Chakraborty}
\affiliation{Department of Chemistry, Indian Institute of Technology Bombay, Powai, Mumbai 400076, India}

\author{B. K. Sahoo}
\email{bijaya@prl.res.in}
\affiliation{Atomic, Molecular and Optical Physics Division, Physical Research Laboratory, Navrangpura, Ahmedabad 380009, India}

\author{J\'{a}n \v{S}imunek}

\author{Jozef  Noga}

\author{Achintya Kumar Dutta}
\email{achintya.kumar.dutta@uniba.sk}
\email{achintya@chem.iitb.ac.in}

\affiliation{ Department of Inorganic Chemistry, Faculty of Natural Sciences, Comenius University, Bratislava Ilkovi\v{c}ova 6, Mlynsk\'a dolina 842 15 Bratislava, Slovakia }

\begin{abstract}
We present accurate values of the electric dipole polarizabilities ($\alpha$) of the ground and excited states of Sr using four-component relativistic coupled-cluster (RCC) theory within the linear-response (LR) and equation-of-motion (EOM) frameworks. We first investigated accuracy of the ground-state polarizability using the LR-RCC method in both spherical and Cartesian coordinate representations. Upon obtaining consistent results for the ground state, we determine the polarizabilities of the excited states using the EOM-RCC method within the finite-field approach. The uncertainties in our results are assessed by examining the completeness of the basis set, higher-order relativistic corrections, and contributions from higher-level excitation configurations beyond the truncation in our RCC method. We also discuss various procedures for extrapolating contributions to the complete basis set limit and examine their respective limitations. Our calculated ground-state polarizability is found to be inconsistent with the latest atomic calculation, but agrees with other calculations and experimental value. Using the differential $\alpha$ value from the EOM-RCC method, we estimate the fractional black-body-radiation-induced frequency shift of the Sr $^{1}$S$_{0}$$\rightarrow$$^{3}$P$_{0}$ clock transition to be $-5.40\times10^{-15}$.
\end{abstract}

\maketitle
\section{Introduction}

Precise electric dipole polarizabilities ($\alpha$) are essential for a wide range of high-precision experiments in atomic and molecular physics \cite{Mitroy_2010, doi:10.1142/2962}. In particular, the rapid development of optical atomic clocks has heightened the need for highly accurate polarizability calculations, as reliable values of $\alpha$ are crucial for evaluating the blackbody-radiation (BBR) shift, which constitutes one of the major sources of systematic uncertainty in optical frequency standards \cite{PhysRevA.74.020502, PhysRevLett.107.143006}. Because precise experimental measurements of $\alpha$ are available only for a limited number of atomic systems, the accuracy of BBR-shift evaluations for clock transitions depends to a large extent on the reliability of theoretical calculations. This makes robust and systematically improvable many-body methods particularly important for determining the polarizabilities of atomic states relevant to optical clocks. As optical clocks approach and surpass the $10^{-18}$ level of fractional frequency uncertainty, even relatively small improvements in the accuracy of calculated polarizabilities can yield meaningful reductions in the uncertainty of BBR shifts, thereby enhancing the performance and reliability of next-generation frequency standards.\\

Extensive theoretical and experimental efforts have therefore been devoted to determining $\alpha$ and evaluating BBR shifts in optical clocks based on a variety of atomic \cite{PhysRevA.90.042505, Dzuba_2019, PhysRevA.93.043420, PhysRevA.109.063111, Sahoo2017, PhysRevLett.109.230802, PhysRevA.98.012513} and ionic \cite{PhysRevA.106.032807, PhysRevA.80.062506, PhysRevA.83.030503, Sahoo2017, PhysRevA.85.012506} systems. For heavy atoms and ions in particular, accurate polarizability calculations require a careful and balanced treatment of electron correlation, higher-order excitations, and relativistic effects, together with QED corrections when their contributions become relevant. Reliable predictions consequently require sophisticated many-body methods capable of incorporating these effects systematically and controllably.\\

Among the current atomic optical-clock candidates, Sr is one of the most extensively studied systems, with a reported fractional frequency uncertainty of approximately $2\times10^{-18}$ \cite{Nicholson2015}. This exceptional level of precision has motivated numerous theoretical investigations of the polarizabilities of the states involved in its clock transition. Several theoretical approaches have been employed, including the relativistic coupled-cluster (RCC) method \cite{BKSahoo1, BKSahoo2, PhysRevA.113.032817, PhysRevA.92.052515}, configuration interaction (CI) combined with the all-order method \cite{PhysRevLett.107.143006, PhysRevA.87.012509, PhysRevA.90.052715}, CI combined with many-body perturbation theory (CI + MBPT) \cite{PhysRevA.78.032508}, and semi-empirical core-polarization methods (CICP) \cite{MitroyZhang2010, PhysRevA.88.022511}. Among these approaches, the RCC framework offers a systematic and balanced treatment of relativistic and electron-correlation effects and is therefore particularly well suited for high-precision polarizability calculations.\\

Despite these extensive theoretical efforts, considerable discrepancies remain among the reported values of the Sr ground-state polarizability. The experimental value is reported with large uncertainty \cite{Schwartz1974}, while theoretical calculations yield a wide range of values \cite{PhysRevA.92.052515, Guo_2010, PhysRevA.78.032508, PhysRevA.74.020502, PhysRevA.87.012509, Mitroy10082010}, generally lying above the central experimental value. More recently, an RCC calculation reported its value to be below the central value of the experimental determination \cite{PhysRevA.113.032817}. This result is particularly noteworthy because the calculation employed an RCC implementation that explicitly exploits the spherical symmetry of atomic systems (hereafter referred to as an atomic code), whereas several earlier calculations were performed using implementations based on symmetry treatments commonly adopted in molecular calculations and formulated in Cartesian coordinates (hereafter referred to as molecular codes). The significant spread among the reported theoretical values thus calls for a systematic investigation of the possible sources of these discrepancies. The situation is even less conclusive for the metastable $^{3}$P$_{0}$ state, for which no experimental measurement of $\alpha$ is currently available, and the existing theoretical predictions also exhibit a considerable spread. These issues provide strong motivation for a comprehensive, high-accuracy study of the polarizabilities of Sr states relevant to optical-clock applications.\\

In this work, we investigate the aforementioned discrepancies by calculating the polarizabilities of the ground and excited states of Sr using different approaches implemented in both atomic and molecular RCC codes. Using atomic codes, we have employed the linear-response (LR) approach using both the ordinary and normal RCC formulations. Using the molecular code, we have calculated the ground-state polarizability using both the LR and finite-field (FF) approaches within the RCC framework. A systematic comparison of the results obtained using these different formulations and implementations allows us to assess their consistency and identify the sources of the discrepancies among the previously reported values. On this basis, we establish a reliable value of the Sr ground-state polarizability. We further have investigated polarizabilities of the metastable states using the FF approach within the LR and equation-of-motion (EOM) RCC theory frameworks implemented in the molecular codes, for which accurate theoretical determinations are particularly important given the lack of experimental values. 

\section{Theory}

The presence of an external electric field inherently mixes states of opposite parity and, therefore, does not preserve parity as a good quantum number. Consequently, it is particularly well suited to molecular implementations formulated in Cartesian coordinates, where such symmetry restrictions are not explicitly imposed. In the FF approach, $\alpha$ is obtained from the second derivative of the atomic energy with respect to an externally applied electric field, which is introduced explicitly into the Hamiltonian. Thus, in this approach the energy associated with a specific state of an atom, when subjected to the influence of an external weak electric field with strength $\textit{\textbf{F}}$, can be described as \cite{MANAKOV1986320,doi:10.1142/2962}
\begin{equation}
\label{eq:2}
E(|\textit{\textbf{F}}|)=E(0)-\frac{\alpha}{2}|\textit{\textbf{F}}|^{2}-\cdots .
\end{equation}
In the above expression, $E(0)$ represents the energy of the state of the atomic Hamiltonian in the absence of any external electric field ($H_0$). By examining Eq. (\ref{eq:2}), it becomes evident that $\alpha$ can be ascertained through the assessment of the second-order derivative of $E(|\textit{\textbf{F}}|)$ due to the interaction Hamiltonian, $H^{'} = -\textit{\textbf{F}}\cdot{\textit{\textbf{D}}}$) with respect to a small electric field of magnitude $\textit{\textbf{F}}$ and induced electric dipole moment $\textit{\textbf{D}}$; i.e.
\begin{equation}
\label{eq:3}
\alpha=-\left(\frac{\delta^{2}E(|\textit{\textbf{F}}|)}{\delta|\textit{\textbf{F}}|\delta|\textit{\textbf{F}}|}\right)_{|\textit{\textbf{F}}|=0} .
\end{equation}
The energy derivative presented in Eq. (\ref{eq:3}) can be computed through both analytical and numerical approaches. The process of calculating the derivative using an analytical approach, also known as the analytical gradient method, requires differentiating the energy functional with respect to the external perturbation (the electric field in the present case), which can be tedious in the RCC method and is neither variational nor Hermitian. In the numerical techniques of the FF approach, the $\alpha$ value is evaluated explicitly by the energy derivatives with respect to $\textit{\textbf{F}}$. It is now clear that the FF approach offers a significant benefit as it effortlessly extends to analyze the characteristics of both the ground and excited states of a system, and the calculations at different field strengths can be performed in parallel.\\

The LR approach provides a fundamentally different route to $\alpha$. 
Atomic systems, in contrast, possess spherical symmetry, and parity remains a good quantum number in the absence of an external perturbation. In an atomic implementation that explicitly exploits these symmetries, introducing a finite external electric field in the FF approach breaks spherical symmetry and mixes states of opposite parity, making the conventional FF formulation less natural for such implementations. The LR approach, on the other hand, can retain the symmetry-adapted atomic framework by treating the electric field as a perturbation and evaluating the corresponding first-order response. It therefore provides a natural and fully first-principles analytic framework for calculating $\alpha$ using atomic codes. To formulate this approach, the wave function of an atomic state due to $\lambda D$, with an arbitrary parameter $\lambda$ having the dimension of electric field, can be expressed as
\begin{eqnarray}
|\Psi \rangle = |\Psi^{(0)} \rangle  + \lambda |\Psi^{(1)} \rangle + \frac{1}{2} \lambda^2  |\Psi^{(2)} \rangle + \cdots ,
\label{eq0}
\end{eqnarray}
where $|\Psi^{(0)} \rangle$ represents unperturbed wave function due to $H_0$ and $|\Psi^{(n)} \rangle$ denotes for the $n^{th}$-order correction due to $D$. This leads to the expectation value of $\lambda D$ with respect to the total wave function as
\begin{eqnarray}
\langle \lambda D \rangle &=&  \frac{\langle \Psi | \lambda D | \Psi \rangle} {\langle \Psi |\Psi \rangle} \nonumber \\
 &=& 2 \frac{  \lambda^2 \langle \Psi^{(0)} | D |\Psi^{(1)} \rangle + \frac{\lambda^4}{24} \langle \Psi^{(1)} | D |\Psi^{(2)} \rangle + \cdots} {\langle \Psi^{(0)} | \Psi^{(0)} \rangle + \frac{1}{2}  \lambda^2 \langle \Psi^{(1)} | \Psi^{(1)} \rangle + \cdots} \ \ \ \ .
 \label{eqld}
\end{eqnarray}
From this expression, it yields 
  \begin{eqnarray}
\alpha = 2 \frac{\langle \Psi^{(0)} | \textbf{D} |\Psi^{(1)} \rangle} {\langle \Psi^{(0)} |\Psi^{(0)} \rangle}  .
\label{eq3}
 \end{eqnarray}
Due to complexity in determining $|\Psi^{(1)} \rangle$, some of the previous atomic calculations use the sum-over-states (SOS) approach for estimating $\alpha$, in which the above expression is given by expanding $|\Psi_0^{(1)} \rangle$ as 
\begin{eqnarray}
\alpha = 2 \sum_{I \ne 0} \frac{|\langle \Psi^{(0)} | \textbf{D} |\Psi_I^{(0)} \rangle|^2} {\langle \Psi^{(0)} |\Psi^{(0)} \rangle (E^{(0)} - E_I^{(0)}) }  ,
\label{eq4}
 \end{eqnarray}
where the summation $I$ runs over intermediate states. The advantage of the SOS approach is that it is possible to use available experimental values of the E1 matrix elements $\langle \Psi_0^{(0)} | \textbf{D} |\Psi_I^{(0)} \rangle$ and energies, computing only $D$ matrix elements for a few dominantly contributing atomic states. However, it misses contributions from core correlations, the continuum, and intermediate states with higher-level excitation configurations. To account for all these contributions on an equal footing, we obtain the first-order wave function by solving the following inhomogeneous equation for the first-order perturbed wave function
\begin{eqnarray}
 (H_0-E^{(0)}) | \Psi^{(1)} \rangle = -D | \Psi^{(1)} \rangle .
\end{eqnarray}

The above LR applies to atomic codes and can also be implemented in molecular codes by treating the external electric field as a time-dependent perturbation arising from a dynamic electric field at frequency $\omega$. As a result, $\alpha$ can be determined as the linear response function of $D$. In the RCC theory framework, this amounts to solving a set of first-order response equations for the perturbed cluster amplitudes at zero field strength \cite{koch1990coupled,10.1063/5.0274450}. The static polarizability is then extracted from the response function in the zero-frequency limit ($\omega \rightarrow 0$). In this approach, an external time-dependent, interaction, $V(t)$, is incorporated into the atomic Hamiltonian, 
\begin{align}
\label{eq:23}
    \hat{H} &= \hat{H}_0 +\hat{V(t)} ,
\end{align}
where
\begin{align}
   \hat{ V(t)} &= \int_{-\infty}^{\infty} V(\omega) e^{(\alpha-i\omega)t}\, d\omega
\end{align}
which is constructed such that it vanishes at $t=-\infty$. For our purpose, we consider $V(t)= D t $ with virtual time $t$. The function $V(\omega)$ is the Fourier transform of $V(t)$, and the small positive quantity $\alpha$ ensures the correct physical boundary condition $V(-\infty)=0$ as described below. Here, the central idea is to determine how the atomic wave function responds to an external time-dependent field. Following the principle of time-dependent perturbation theory, it yields
\begin{equation}
\label{eq:25}
    i\frac{d}{dt} \left| {\Psi } (t)\right\rangle = (\hat{H}_0 + \hat{V}(t)) \left| {\Psi } (t)\right\rangle 
\end{equation}
such that the time-dependent wave function, $\left| {\Psi } (t)\right\rangle$, can be written as \cite{olsen1985linear}
\begin{equation}
\label{eq:26}
    \left| {\Psi } (t)\right\rangle = \left| {\overline{\Psi} } \right\rangle e^{i\epsilon(t)} ,
\end{equation}
where $\epsilon(t)$ is a real phase and $\left| {\overline{\Psi} }_{0}\right\rangle$ is the phase-independent component. This state can be expanded perturbatively as
\begin{equation}
\label{eq:27}
   \left| {\overline{\Psi} } \right\rangle = \left| {\Psi^{(0)} } \right \rangle + \left| {\Psi^{(1)} } \right\rangle + {\left| {\Psi^{(0)} } \right\rangle} + ... .
\end{equation}

Following Eq. (\ref{eqld}), we can determine $\alpha$ by evaluating expectation value of the operator $\hat{A}= D t$ (same as $V(t)$ for the present choice of the problem). Using the time-dependent wave function, the expectation value of $\hat{A}$ can be expressed as a sum of response contributions of increasing order \cite{olsen1985linear} ,
\begin{equation}
\langle \hat A \rangle(t) =  \langle \Psi_0 | \hat A | \Psi_0 \rangle + \langle A \rangle^{(1)}(t) + \langle A \rangle^{(2)}(t) + \cdots,
\label{eq:28}
\end{equation}
where the first-order (linear), second-order (quadratic), etc. response terms are
\begin{align}
\langle A \rangle^{(1)}(t) &= \int_{-\infty}^{\infty}\!\! \langle\langle \hat A; V^{\omega_1} \rangle\rangle_{\omega_1+i\alpha}\, e^{-i(\omega_1+i\alpha)t} \, d\omega_1, \\
\langle A \rangle^{(2)}(t) &= \tfrac{1}{2}\int\!\!\int_{-\infty}^{\infty}\!\! \langle\langle \hat A; V^{\omega_1}; V^{\omega_2} \rangle\rangle_{\omega_1+i\alpha,\,\omega_2+i\alpha} \notag \\
&\quad \times\, e^{-i(\omega_1+\omega_2+2i\alpha)t} \, d\omega_1 \, d\omega_2,
\end{align}
and so on with $\langle\langle \hat A; V^{\omega_1} \rangle\rangle$, $\langle\langle \hat A; V^{\omega_1}; V^{\omega_2} \rangle\rangle$, etc. are the linear, quadratic and higher-order response functions, respectively. From this expansion, it follows that the LR function corresponds to the first-order change in the expectation value of $\hat{A}$. For exact wave functions, the LR function may be written in the sum-over-states form:
\begin{eqnarray}
\left\langle \left\langle \hat{A}; V^{\omega_1}\right\rangle \right\rangle =&\sum\limits_{k}\left[\frac{\left\langle  {{\Psi }_{0}} \right|\hat{A}\left| {{\Psi }_{k}} \right\rangle\left\langle  {{\Psi }_{k}} \right|V^{\omega_1}\left| {{\Psi }_{0}} \right\rangle}{\omega_1 - \omega_k}\right] \nonumber \\
&-\sum\limits_{k} \left[ \frac{\left\langle  {{\Psi }_{0}} \right|V^{\omega_1}\left| {{\Psi }_{k}} \right\rangle\left\langle  {{\Psi }_{k}} \right|\hat{A}\left| {{\Psi }_{0}} \right\rangle}{\omega_1 + \omega_k} \right]
\end{eqnarray}
where $\omega_k$ denotes the excitation energies from the ground state to excited states. Comparing it with Eq. (\ref{eq4}) and setting $\omega=0$, we can obtain the $\alpha$ value. 

\section{Methods for calculations}

Below, we discuss the FF and LR approaches employed in the atomic and molecular codes to evaluate the ground- and excited-state polarizabilities of Sr. Although the same approaches are used in both codes, the implementations differ. These atomic codes with spherical symmetry have been successfully employed to evaluate $\alpha$ of several systems \cite{BKSahoo1, BKSahoo2, BKSahoo3, BKSahoo4, BKSahoo5}. Similarly, the molecular codes used here are available in the BAGH software package \cite{BAGH} which is used to calculate properties of both atoms and molecules. 

\subsection{Approximated Hamiltonian and orbital wave functions}

We begin with $\hat{H}_0$ as the Dirac-Coloumb (DC) Hamiltonian in our calculations, which is defined in atomic units (a.u.) as
\begin{equation}
\label{eq:8}
\hat{H}_0=\sum_i^N\left[c \boldsymbol{\alpha} \cdot \boldsymbol{p}_i + (\beta_i -1)  c^2+V_{n}(r_i)\right]+\sum_{i>j}^N \frac{1}{r_{i j}} .
\end{equation}
In Eq. (\ref{eq:8}), $\boldsymbol{\alpha}$ and $\beta$ are the Dirac matrices, $V_{n}(r)$ represents the nuclear potential defined using the Fermi nuclear charge distribution, and $c$ stands for the speed of light. For the correlation-treated no-pair approximation \cite{PhysRevA.22.348} is used, where the summation in Eq. (\ref{eq:8}) is restricted to the positive energy spectrum only. 
We have also included corrections from the Breit interaction by subsequently adding the following term to the DC Hamiltonian
\begin{equation}
\label{eq:10}
B_{12} =- \sum_{i>j}^N \frac{1}{2 r_{i j}}\left[\boldsymbol{\alpha_i} \cdot \boldsymbol{\alpha_j}+\frac{\left(\boldsymbol{\alpha_i} \cdot \boldsymbol{r_{i j}}\right)\left(\boldsymbol{\alpha_j} \cdot \boldsymbol{r_{i j}}\right)}{r_{i j}^2}\right] .
\end{equation}
We evaluate results by dividing the Breit term into the Gaunt term,
\begin{equation}
\label{eq:11}
G_{12}=- \sum_{i>j}^N \frac{\boldsymbol{\alpha_i} \cdot \boldsymbol{\alpha_j}}{2 r_{i j}} ,
\end{equation}
and the retardation term,
\begin{equation}
\label{eq:12}
B_{12} - G_{12} =-\frac{1}{2 r_{i j}}\left[\frac{\left(\boldsymbol{\alpha_i} \cdot \boldsymbol{r_{i j}}\right)\left(\boldsymbol{\alpha_j} \cdot \boldsymbol{r_{i j}}\right)}{r_{i j}^2}\right] ,
\end{equation}
to verify individual contributions explicitly.

In the atomic code, we incorporate vacuum-polarization contributions via the local Uehling potential, 
\begin{eqnarray}
V_U &=& - \sum_i \frac{2 }{3 } \frac{\alpha_e^2}{r_i} \int_0^{\infty} dx \, x \, \rho_n(x) \, \int_1^\infty dt\, \sqrt{t^2-1} \nonumber \\
 && \times \left(\frac{1}{t^3}+\frac{1}{2t^5}\right) \left [ e^{-2ct|r_i-x|} - e^{-2ct|r_i+x|}  \right ], \ \ \ \ \
\end{eqnarray}
where $\alpha_e$ is the fine-structure constant and $\rho_n(r)$ is the nuclear density normalized to atomic number $Z$.

Since $H_0$ has both one- and two-body terms, its solution is first obtained in the mean-field approach using the Dirac-Hartree-Fock (DHF) method by defining $H_0 = H_{DHF} + V_{res}$, where $H_{DHF}$ is the DHF Hamiltonian and $V_{res}$ is the residual interaction that is neglected at the DHF method. Using the DHF Hamiltonian, we obtain the single-particle wave function of the $k^{th}$ orbital as
\begin{equation}
\phi_{k}(r)=\frac{1}{r}
\begin{bmatrix}
 P_k(r) & \chi_{\kappa,j,m_j}(\theta, \phi) \\
 i Q_k (r) & \chi_{-\kappa,j,m_j}(\theta, \phi) \\
\end{bmatrix},
\end{equation}
where $P(r)$ and $Q(r)$ are the large and small components of the radial wave functions, respectively, $\chi_{\kappa,j,m_j}(\theta, \phi)$ denotes the angular component with relativistic quantum number $\kappa$, total angular momentum $j$ and its component $m_j$. These angular factors are dealt with explicitly in our atomic codes whereas orbitals for the molecular codes are obtained by taking the linear combinations of atomic orbitals. To generate all the orbitals, we have used Gaussian type orbitals as basis functions for the radial components satisfying even-tempering conditions, and the kinetic balance condition between the $P(r)$ and $Q(r)$ components.

Since the atomic codes are spherical symmetry, we compute reduced matrix elements rather than actual matrix elements to minimize computational cost. In terms of reduced matrix elements, the actual matrix elements of the one-body and two-body operators are typically expressed as \cite{Edmond}
\begin{eqnarray}
\langle J', M_{J'} | O_1 | JM_J \rangle &=& \sum_{k,q} (-1)^{J'-M_{J'}} \begin{pmatrix}
 J' & k & J \\
 -M_{J'} & q & M_J \\
\end{pmatrix} \nonumber \\ && \times \langle J' || O_1^{(k)} || J \rangle
\end{eqnarray}
and 
\begin{eqnarray}
\langle J_a, M_{J_a}, J_b M_{J_b} | O_1 \otimes O_2 | J_c M_{J_c}, J_d M_{J_d} \rangle = \sum_{k,q} R_{abcd}^{(k)} 
 \nonumber \\ \times \begin{pmatrix}
 J_a & k & J_c \\
 -M_{J_a} & q & M_{J_c} \\
 \end{pmatrix} 
 \begin{pmatrix}
 J_b & k & J_d \\
 -M_{J_b} & -q & M_{J_d} \\
 \end{pmatrix} \nonumber \\
 \times (-1)^{J_a-m_{J_a} + J_b - M_{J_b}+k-q}
\end{eqnarray}
with the one-body and two-body reduced matrix elements $\langle J' || O_1^{(k)} || J \rangle$ and $R_{abcd}^{(k)} =(-1)^k \langle J_a ||O_1^{(k)}|| J_c \rangle \langle J_b ||O_2^{(k)}|| J_d \rangle$, respectively. In these expressions, $k$s are called as multiples with components $q=-k, -k+1, \cdots, k-1, k$. In a nutshell, our molecular codes deal with the actual matrix elements, while our atomic codes deal with the reduced matrix elements with respect to the angular momentum of the orbitals. Computational costs in the molecular and atomic codes are optimized through the group symmetry properties and use of reduced matrix elements, respectively.

\subsection{RCC method for the ground state wave function}

To account for the correlation effect from $V_{res}$, the wave function of the ground state of an atomic or a molecular system in the RCC theory formalism is expressed as
\begin{equation}
\label{eq:4}
|\Psi_{0}\rangle=e^{\hat{T}}|\Phi_{0}\rangle ,
\end{equation}
where $|\Phi_{0} \rangle$ denotes the reference determinant constructed using the DHF orbitals $\phi(r)$, and $\hat{T}$ is the RCC operator that generates different levels of excitations from $|\Phi_{0}\rangle$. These excitation levels, for an $N$ number of electron system, are denoted by subscripts as
\begin{equation}
\label{eq:5}
\hat{T}=\hat{T}_{1}+\hat{T}_{2}+ \cdots + \hat{T}_{N} ,
\end{equation}
where any general $n$-tuple excitation RCC operator can be expressed as
\begin{equation}
\label{eq:6}
\hat{T_{N}}=\left(\frac{1}{N!}\right)^{2}\sum_{ij...ab...}{t_{ij...}^{ab...}}a_{a}^{\dagger}a_{b}^{\dagger}...\hspace{0.2cm}...a_{j}a_{i}.
\end{equation}
In Eq. (\ref{eq:6}), $t_{ij...}^{ab...}$ are the cluster amplitudes, indices $\left(i\hspace{0.1cm},j \hspace{0.1cm},k\hspace{0.1cm}...\right)$ represent occupied spinors and virtual spinors are shown by $\left(a\hspace{0.1cm},b\hspace{0.1cm},c\hspace{0.1cm}...\right)$. It is worth mentioning here that the above RCC amplitudes are obtained as the actual values in both molecular and atomic codes, {\it albeit} the reduced matrix elements are used for their evaluation in the atomic codes. The commonly used singles-doubles approximation in the RCC (RCCSD) method is obtained by truncating the cluster operator to include only one- and two-body excitations. The RCC amplitudes are obtained by solving a set of simultaneous nonlinear equations
\begin{equation}
\label{eq:7}
\langle \Phi_{ij...}^{ab...}|\Bar{H}|\Phi_{0}\rangle=0 ,
\end{equation}
where $ | \Phi_{ij..}^{ab..} \rangle$ are the excited determinants and
$\Bar{H}=e^{-\hat{T}}\hat{H}_0e^{\hat{T}}$ is the similarity transformed Hamiltonian. 

The ground state energy expression in the RCC theory is given as
\begin{equation}
\label{eq:13}
\langle \Phi_{0}|\Bar{H}|\Phi_{0}\rangle=E .
\end{equation}
It should be noted that the RCC expressions for the determination or wave function amplitudes and $E$ are same in the absence and presence of interaction Hamiltonian. Therefore, no extension of the RCC method is required to estimate $\alpha$ value of the ground state in the FF-RCC method. Again, the RCCSD approximation in the RCC theory frequently falls short in delivering precise quantitative accuracy, prompting one to incorporate additional higher-level excitations in the form of triples (RCCSDT), quadruples (RCCSDTQ), pentuples (RCCSDTQP), etc. to achieve systematically accurate results at the expense of significantly higher computational cost. Due to limitations in computational resources with us, we have carried out calculations primarily using the RCCSD method and then included triples perturbatively (RCCSD(T) and CC3 or its higher variants). Wherever possible, we have considered full triple excitations through the RCCSDT method and along with it we have considered quadruple contributions perturbatively (RCCSDT(Q)).

\subsection{LR approach of the atomic code}

Since atomic code explicitly obeys spherical symmetric properties, the DHF orbitals generated using this code treat parity as a good quantum number. As a result, the FF approach cannot be employed using the atomic code and the LR approach is the ultimate choice to determine $\alpha$ using this code. It means that the DHF value of $\alpha$ in the LR approach misses the orbital relaxation effects that are naturally present in the DHF value of the FF approach. In the LR approach, these orbital relaxation effects in the evaluation of $\alpha$ at the DHF level can be accounted through the random phase approximation (RPA) using a correlation operator $\Omega_{RPA}$ as
\begin{eqnarray}
\alpha = 2\langle \Phi_0 | D \Omega_{RPA} | \Phi_0 \rangle , 
\end{eqnarray}
where the perturbed wave operator in the RPA is given by \cite{BKSahoo6}
\begin{eqnarray}
\Omega_{RPA} &=& \sum_{k=0}^{\infty} \sum_{a,i} \Omega_{i,a}^{(k)} =  \sum_{a,i} \Omega_{i,a}^{(0)}  \nonumber \\
     &+& \sum_{k=1}^{\infty} \sum_{ab,ip} \left ( \frac{\langle \Phi_{ij}^{ab}| H_{at} |\Phi_j^b \rangle \langle \Phi_j^b |\Omega_{b,j}^{(k-1)} | \Phi_0 \rangle}{\epsilon_i - \epsilon_a} \right. \nonumber \\
     &+& \left.  \frac{\langle \Phi_i^a| \Omega_{b,j}^{(k-1)\dagger} |\Phi_{ij}^{ab} \rangle \langle \Phi_{ij}^{ab} | H_{at} | \Phi_0 \rangle}{\epsilon_i - \epsilon_a}  \right ).
\end{eqnarray}
Here we assume the initial guess of the $\Omega_{RPA}$ operator to be $\Omega_{i,a}^{(0)} = \frac{\langle \Phi_i^a | D | \Phi_0 \rangle}{\epsilon_i - \epsilon_a} a_a^{\dagger} a_i $ and the final solution is obtained through iteration procedure.

To obtain the first-order wave function of the ground state in the LR-RCC method, we write   
\begin{eqnarray}
T = T^{(0)} + \lambda T^{(1)} + \frac{1}{2} \lambda^2 T^{(2)} + \cdots ,
\label{eq10}
\end{eqnarray}
where superscripts denote the order of the $D$ operator in this case. Then, the unperturbed and first-order wave functions are given by
\begin{eqnarray}
  |\Psi_0^{(0)} \rangle = e^{T^{(0)}} | \Phi_0 \rangle 
\end{eqnarray}
and
\begin{eqnarray}
  \Psi_0^{(1)} \rangle = e^{T^{(0)}} T^{(1)}| \Phi_0 \rangle ,
\end{eqnarray}
respectively. It follows \cite{BKSahoo7}
\begin{eqnarray}
\alpha &=& 2 \frac{\langle \Phi_0 | \bar{D} T^{(1)} | \Phi_0 \rangle} {\langle \Phi_0 | e^{T^{(0)\dagger}} e^{T^{(0)}} | \Phi_0 \rangle} \nonumber \\
      &=& 2 \langle \Phi_0 | \bar{D} T^{(1)} | \Phi_0 \rangle_l  ,
\label{eq12}
\end{eqnarray}
where $\bar{D}=e^{T^{(0)\dagger}} D e^{T^{(0)}}$ and subscript $l$ denotes for the linked terms. The non-terminating expression of $\bar{D}$ is evaluated self-consistently by dividing it into effective one-body and two-body terms.

In Eq. (\ref{eq12}), it can be argued that the normalization factor  may not be exactly canceled out with the disconnected part of the numerator when the RCC method is approximated. In order to avoid any ambiguity in accounting for the contribution from the normalization of the wave function, we employ the normal RCC (NRCC) method, proposed in Ref. \cite{BKSahoo4, BKSahoo5}, to estimate the $\alpha$ value using the atomic codes. In this method, the ket of the state is the same as in the RCC method, but the bra state is given by \cite{Monkhorst, koch1990coupled}
\begin{eqnarray}
\langle \tilde{\Psi_0 } | = \langle \Phi_0 | (1+ \Lambda ) e^{-T}, 
\end{eqnarray}
where $\Lambda$ is a de-excitation operator similar to $T^{\dagger}$ but with different amplitudes such that
\begin{eqnarray}
\langle \tilde{\Psi_0} | \Psi_0 \rangle = 1
\end{eqnarray}
and 
\begin{eqnarray}
\langle \Phi_0 | \Lambda H \Psi_0 \rangle = 0.
\end{eqnarray}
The last equation acts as a precondition to ensure that both $\langle \tilde{\Psi_0} |$ and $\langle \Psi_0 |$ describe the same state of energy $E$. 

The next step is to perturb the $\Lambda$ operator, i.e.
\begin{eqnarray}
 \Lambda = \Lambda^{(0)} + \lambda \Lambda^{(1)} + \frac{1}{2} \lambda^2 \Lambda^{(2)} + \cdots .
\end{eqnarray}
The amplitude determining equations for $\Lambda^{(0)}$ and $\Lambda^{(1)}$ are given by
\begin{eqnarray}
\langle \Phi_0 | \Lambda^{(0)} \Bar{H} + \Bar{H} | \Phi_ {ij...}^{ab...} \rangle =0 
\end{eqnarray}
and
\begin{eqnarray}
\langle \Phi_0 | \Lambda^{(1)} \Bar{H} + (\Lambda^{(0)} + 1) \left [  \Bar{H} T^{(1)} + \Bar{D} \right ] | \Phi_ {ij...}^{ab...} \rangle =0 ,
\end{eqnarray}
respectively.

 The $\alpha$ evaluating expression in the RNCC method is given by
\begin{eqnarray}
\alpha =  \langle \Phi_0 | (D e^{(0)})_l T^{(1)} | \Phi_0 \rangle_l + \langle \Phi_0 |  \Lambda^{(1)} (D e^{(0)})_l | \Phi_0 \rangle_l .
\label{eq20}
\end{eqnarray}  

At this juncture, we would like to mention that the RPA effects are embedded within the perturbative RCC theory framework through the construction of the singly excitations of the $T^{(1)}$ operator, along with many other correlation effects. Thus, a good agreement between the DHF value of $\alpha$ from the FF approach of the molecular code and the RPA value of $\alpha$ from the atomic code would be able to validate correct implementation of both the codes.

\begin{table*}[!t]
\centering
\caption{The ground- and low-lying excited-state $\alpha$ values (in a.u.) of Sr reported over the past two decades using various many-body methods, together with a comparison with the available experimental value for the ground state.}
\label{tab:full_sr_results_new}
\begin{tabular}{@{}l r r r r l c@{}}
\hline\hline
\text{Source}& \multicolumn{4}{c}{\text{$\alpha$ value (a.u.)}} & Method/ & Year of  \\
\cline{2-5} \\
 & \multicolumn{1}{c}{$^1S_0$} & \multicolumn{1}{c}{$^3P_0$} &
\multicolumn{1}{c}{$^3P_1$ ($|M_J|=1$)} & \multicolumn{1}{c}{$^3P_2$ ($|M_J|=0$)} & technique & reporting \\
\hline \\
\text{Previous calculations} & & & & \\
Ref. \cite{PhysRevA.113.032817} & 176.50(2.65) & & & & PRCC(T) & 2026 \\
Ref. \cite{wu2019dynamic} & 202.02 & & & & DFCP+RCI & 2019 \\
Ref. \cite{PhysRevA.92.052515} & 199.7 & 444.1 & 480.9 & 543.9 & EOM-RCC & 2015 \\
Ref. \cite{Chattopadhyay2014} & 190.82 & & & & PRCC & 2014 \\
Ref. \cite{PhysRevA.87.012509} & 194.4 & 441.9 & & & CI+All-order & 2013 \\
Ref. \cite{SinghSahooDas2013} & 186.98(85) & & & & PRCC & 2013 \\
Ref. \cite{PhysRevA.88.022511} & 197.8 & & & & CICP & 2013 \\
Ref. \cite{Guo_2010} & 193(13) & 410(28) & & & CICP & 2010 \\
Ref. \cite{Mitroy10082010} & 204.5 & 497.0 & & & CICP & 2010 \\
Ref. \cite{PhysRevA.78.032508} & 197.2 & 457.0 & 498.8 & & CI+MBPT & 2008 \\
Ref. \cite{BKSahoo2} & 199.71(7.28) & & & & PRCC & 2008 \\
Ref. \cite{PhysRevA.74.020502} & 197.2(2) & 458.3(3.6) & & & CI+MBPT & 2006 \\
Ref. \cite{porsev2006high} & 202.0 & & & & CI+MBPT & 2006 \\
Ref. \cite{LimSchwerdtfeger2004} & 198.85 & & & & DK-CCSD(T) & 2004 \\
& & & & \\
\text{Experiment} \cite{Schwartz1974} & 186(15) & & & & E-H Balance & 1974 \\
\hline \hline
\end{tabular}
\end{table*}

\subsection{LR approach of the molecular code}

In our molecular code, the LR approach is implemented for the time dependent interactions by expressing
\begin{eqnarray}
        \left| {\Psi }_{0} (t)\right\rangle = e^{\hat{T}(t)}\left| {\Phi }_{0} (t)\right\rangle e^{i\epsilon(t)} 
\end{eqnarray}
and
\begin{eqnarray}
    \left\langle \tilde{\Psi}_0 (t)\right| &=&  \langle \Phi_0 (t) | (1+ \hat{\Lambda}(t) ) e^{-T(t)} e^{-i\epsilon(t)} \nonumber \\ 
    &=& \{ \left\langle  {{\Phi }_{0}} \right| + \sum\limits_{\mu} \lambda_\mu (t) \left\langle  {\mu } \right| e^{-\hat{T}(t)} \} e^{-i\epsilon(t)}
\end{eqnarray}
where the exponential factor carries the time-dependent phase. Using Eq. (\ref{eq:25}), one obtains expressions for the time derivatives of the amplitudes $t_\mu$ and $\lambda_\mu$:
\begin{eqnarray}
    \frac{dt_\mu}{dt} = -i \left\langle  {\mu}\right|e^{-{\hat{T}(t)}}(\hat{H}_0 + \hat{V}(t))e^{\hat{T}(t)}\left| {{\Phi }_{0}} \right\rangle 
\end{eqnarray}
and
\begin{eqnarray}
    \frac{d\lambda_\mu}{dt} = i \left\langle  { \bar{\tilde{\Psi}}_0 (t)}\right|\left[\hat{H}_0 + \hat{V}(t), \tau_\mu\right]\left| {\bar{{\Psi }}_{0}(t)} \right\rangle
\end{eqnarray}
with
\begin{eqnarray}
        \left\langle  {\bar{\tilde{\Psi}}_0(t)}\right|= \left\langle  {{\Phi }_{0}} \right| + \sum\limits_{\mu} \lambda_\mu (t) \left\langle  {\mu } \right| e^{-\hat{T}(t)}
\end{eqnarray}
and
\begin{eqnarray}
    \left| {\widetilde{\Psi }}_{cc} (t)\right\rangle = e^{\hat{T}(t)}\left| {\Phi }_{0} (t)\right\rangle .
\end{eqnarray}

In terms of perturbation expression, we can write
\begin{align}
    t_\mu(t) &= t_{\mu}^{(0)} + t_{\mu}^{(1)}+t_{\mu}^{(2)}+ \cdots 
\end{align}
and
\begin{align}
    \lambda_\mu(t) &= \lambda_{\mu}^{(0)} + \lambda_{\mu}^{(1)} + \lambda_{\mu}^{(2)}+ \cdots .
\end{align}
Again using Fourier transformation, one obtains
\begin{align}
    t_\mu(t) &= t_{\mu}^{(0)} + \int_{-\infty}^{\infty}X_{\mu}^{(1)} e^{(-i\omega_1 + \alpha)t} \, d\omega_1 + \cdots 
\end{align}
and
\begin{align}
    \lambda_\mu(t) &= \lambda_{\mu}^{(0)} + \int_{-\infty}^{\infty}Y_{\mu}^{(1)} e^{(-i\omega_1 + \alpha)t} \, d\omega_1 + \cdots ,
\end{align}
where $X_{\mu}^{(1)}$ and $Y_{\mu}^{(1)}$ denote the Fourier-transformed first-order amplitudes. The right-hand perturbed amplitudes satisfy
\begin{align}
    X_{\mu}^{(1)}(\omega_1+i\alpha) &= \sum_{\nu} \left[-\mathbf{A}+(\omega_1 + i\alpha)\mathbf{I}\right]^{-1}_{\mu\nu}\xi_{\nu}^{(1)}(\omega_1) .
\end{align} 
Here, $\mathbf{I}$ is the identity matrix and $\mathbf{A}$ is the RCC Jacobian,
\begin{equation}
    A_{\mu\nu} = \left\langle  {\mu} \right| \left[ \Bar{H},\tau_\nu\right] \left| {{\Phi }_{0}} \right\rangle
\end{equation}
while
\begin{equation}
    \xi_{\nu}^{(1)}(\omega_1) = \left\langle  {\nu} \right|\overline{V}{^{(\omega_1)}}\left| {{\Phi }_{0}} \right\rangle .
\end{equation}
Similarly, the left-hand perturbed amplitudes follow
\begin{eqnarray}
  Y_{\mu }^{(1)}({{\omega }_{1}}+i\alpha )\text{ } &=&  \sum\limits_{\gamma }{{{F}_{\nu \gamma }}} X_{\gamma }^{(1)} ({{\omega }_{1}}+i\alpha )\} \nonumber \\ && \times  \left\{ \mathbf{A}+({{\omega }_{1}}+i\alpha )\mathbf{I} \right\}_{\nu \gamma }^{-1} \nonumber \\ && -\sum\limits_{\nu }{\{\eta _{\nu }^{(1)}}({{\omega }_{1}}), \ \ \ \ \ \ \ \ 
\end{eqnarray}
where $\eta^{(1)}$ and the matrix $\mathbf{F}$ are defined as
\begin{align}
    &\eta_{\nu}^{(1)}(\omega_1) = \left\langle  {(1 + \hat{\Lambda})} \right|[\,\overline{V}{^{(\omega_1)},\tau_\nu}]\left| {{\Phi }_{0}} \right\rangle
\end{align}
and
\begin{align}
    &F_{\nu\gamma} = \left\langle  {(1 + \hat{\Lambda})} \right|[[\Bar{H}, \tau_\nu],\tau_\gamma ]]\left| {{\Phi }_{0}} \right\rangle .
\end{align}

\begin{figure}[t]
\centering
\includegraphics[width=\columnwidth, height=6cm]{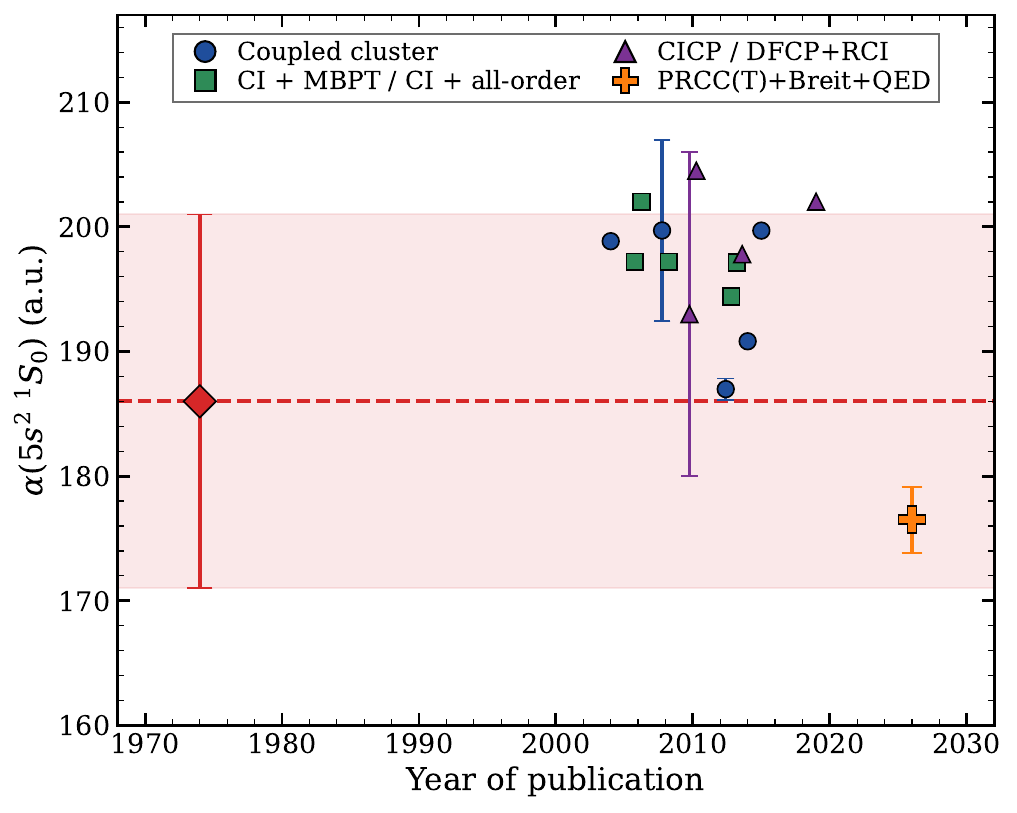}
\caption{ 
Reported theoretical and experimental ground-state $\alpha$ values of Sr over the past five decades. The theoretical values (blue circles) reported between 2004 and 2015 span approximately $187$--$202$~a.u. The experimental value reported by Schwartz, Miller, and Bederson \cite{Schwartz1974} (red diamond, with the dashed line indicating $186$~a.u.) has an uncertainty of $\pm15$~a.u., which has remained essentially unchanged since its original measurement. As can be seen, the theoretical values span nearly the entire experimental uncertainty range, with different calculations reporting high-precision results despite their substantial spread.
} 
  \label{fig:sr-pol-history}
\end{figure}

The expectation value, for the evaluation of $\alpha$, can now be obtained using the formula
\begin{align}
\label{eq:47}
    \left\langle \left\langle \textbf{A};\textbf{B} \right\rangle \right\rangle &= \frac{1}{2} \hat{P}(A,B) \left[ \left\langle \Phi_0 \left| [Y_{\omega_1}^{B}, \bar{A}] \right| \Phi_0 \right\rangle \right. \notag \\
    &\quad + \left. \left\langle \Phi_0 \left|(1+\hat{\Lambda}) [\bar{A},X_{\omega_1}^{B}] \right| \Phi_0 \right\rangle \right] ,
\end{align}
where $B=V^{\omega_1}$ and $\hat{P}(A,B)$ exchanges the operators $A$ and $B$ while also performing complex conjugation. The quantities $X_{\omega_1}^{B}$ and $Y_{\omega_1}^{B}$ are the perturbed right and left RCC amplitudes associated with the operator $B$.

An alternative formulation avoids solving for $Y_{\omega_1}^{B}$ by introducing an additional set of amplitudes $X_{-\omega_1}^{B}$:
\begin{equation}
\begin{aligned}
\label{eq:48}
\langle\langle \mathbf{A}; \mathbf{B} \rangle\rangle_{\omega_{1}} &=  
\frac{1}{2} \hat{C}^{\pm \omega_{1}} \hat{P} [A(-\omega_{1}), B(+\omega_{1})] 
\Big\langle \Phi_0 \Big| (1 + \hat{\Lambda}) \\
&\quad \times \Big( [\bar{A}, \hat{X}^{B}_{\omega_{1}}] 
+ \frac{1}{2} [[\bar{H}, \hat{X}^{A}_{-\omega_{1}}], \hat{X}^{B}_{\omega_{1}}] \Big) 
\Big| \Phi_0 \Big\rangle
\end{aligned}
\end{equation}
Here, $\hat{C}$ performs complex conjugation and simultaneously reverses the sign of the field frequency, while $\hat{P}$ ensures operator symmetry. Both formulations, Eqs.~(\ref{eq:47}) and ~(\ref{eq:48}), require comparable computational effort. In the present LR-RCC implementation\cite{10.1063/5.0274450}, we employ the approach of Eq.~(\ref{eq:47}), solving both $X_{\omega_1}^{B}$ and $Y_{\omega_1}^{B}$ amplitudes.

\begin{table}[t]
\centering
\caption{Calculated ground state $\alpha$ value (in a.u.) of Sr at different level of approximations using atomic code. The recommended value from the analysis of contributions from the DC Hamiltonian and higher-order corrections is quoted at the bottom of the table.}
\begin{tabular}{lcc}
\hline \hline \\
Code & Method & $\alpha$ (a.u)\\
\hline \\      
\multicolumn{3}{c}{DC Hamiltonian} \\
   & & \\
Atomic & DHF & 156.91 \\
Atomic & RPA & 232.55 \\
Atomic & RCCSD & 197.24  \\
Atomic & RCCSD(T) & 195.25 \\
Atomic & RNCCSD & 185.55    \\
\hline \\
\multicolumn{3}{c}{Higher-order contributions} \\
  & & \\
Atomic & $\Delta$Breit  & $-0.06$ \\
Atomic & $\Delta$VP  & $-0.01$ \\
 \hline \\
\multicolumn{2}{c}{Recommended} & 195(5)  \\
\hline \hline \\
\end{tabular}
\label{tab1}
\end{table} 

\subsection{EOM-RCC method for the excited states}

As mentioned before, the EOM-RCC method is employed to obtain the excited state wave functions following the works of Refs. \cite{ROWE1968, MUKHERJEE1979325,10.1063/1.464746}. In this method, the $k^{th}$ excited state is expressed as
\begin{equation}
\label{eq:15}
|\Psi_{k}\rangle=\hat{R}_{k}e^{\hat{T}}|\Phi_{0}\rangle ,
\end{equation}
where $\hat{R}_{k}$ is a linear excitation operator with a form of
\begin{equation}
\label{eq:16}
\hat{R}_{k}= r_{0}+\sum_{i,a} r_{i}^{a}\;\hat{a}_{a}^{\dagger}\hat{a}_{i}+\sum_{i<j\atop a < b} r_{i j}^{ab}\;\hat{a}_a^{\dagger}\hat{a}_b^{\dagger} \hat{a}_j \hat{a}_{i}+\cdots .
\end{equation}
In Eq. (\ref{eq:16}), $r_{i}^{a}$, $r_{ij}^{ab}$ $\cdots$ are the amplitudes corresponding to singly, doubly, and higher excited configurations and $r_{0}$ is a constant denoting the overlap with the ground state. The Dirac equation for the $k^{th}$ excited state in the EOM-RCC picture can be written as 
\begin{equation}
\label{eq:17}
\hat{H}_0 \hat{R}_{k}e^{\hat{T}}|\Phi_{0}\rangle=E_{k}\hat{R}_{k}e^{\hat{T}}|\Phi_{0}\rangle .
\end{equation}
Using the commutative property of the $\hat{R}_{k}$ and $\hat{T}$ operators, one arrives at
\begin{equation}
\label{eq:18}
\Bar{H}\hat{R}_{k}|\Phi_{0}\rangle=E_{k}\hat{R}_{k}|\Phi_{0}\rangle
\end{equation}
and
\begin{equation}
\label{eq:19}
\hat{R}_{k}\Bar{H}|\Phi_{0}\rangle=E_{0}\hat{R}_{k}|\Phi_{0}\rangle .
\end{equation}
Taking difference between Eq. (\ref{eq:18}) and Eq. (\ref{eq:19}), we get
\begin{equation}
\label{eq:20}
[\Bar{H} , \hat{R}_{k}]|\Phi_{0}\rangle=\omega_{k}\hat{R}_{k}|\Phi_{0}\rangle .
\end{equation}
Here $\omega_{k}=E_{k}-E_{0}$ is the excitation energy. Similar to the RCCSD approximation, the EOM-CC equations are generally truncated at the singles and doubles approximation, resulting in the EOM-RCCSD method. Davidson's iterative diagonalization method \cite{Hirao1982} is typically employed to solve the EOM-RCC equations. It scales as iterative $O(N^6)$ and $O(N^4)$ storage requirements.  

\begin{table}[t]
\centering
\caption{Contributions (in a.u.) from individual RCCSD and RNCCSD terms to the ground state $\alpha$ (in a.u.) of Sr using atomic code.}
\begin{tabular}{lr|lr}
\hline \hline \\
RCC term & Value & RNCC term & Value \\
\hline \\      
$D T_1^{(1)}$ & 114.41  & $D T_1^{(1)}$ &  114.41 \\
$T_1^{(1)\dagger} D$ & 114.41 &  $\Lambda_1^{(1)\dagger} D$ & 83.77 \\
$T_1^{(0)\dagger} D T_1^{(1)}$ & $-7.24$ & $\Lambda_1^{(0)\dagger} D T_1^{(1)}$ & $-3.57$\\
$T_1^{(1)\dagger} D T_1^{(0)}$ & $-7.24$ & $\Lambda_1^{(1)\dagger} D T_1^{(0)}$ & $-5.35$ \\
$T_2^{(0)\dagger} D T_1^{(1)}$ & $-14.84$ &  $\Lambda_2^{(0)\dagger} D T_1^{(1)}$ &  0.0 \\
$T_1^{(1)\dagger} D T_2^{(0)}$ & $-14.84$ &  $\Lambda_1^{(1)\dagger} D T_2^{(0)}$ & $-10.86$ \\
$T_1^{(0)\dagger} D T_2^{(1)}$ & 0.77 &  $\Lambda_1^{(0)\dagger} D T_2^{(1)}$ & 0.30 \\
$T_2^{(1)\dagger} D T_1^{(0)}$ & 0.77 &  $\Lambda_2^{(1)\dagger} D T_1^{(0)}$ & 0.0 \\
$T_2^{(0)\dagger} D T_2^{(1)}$ & 8.70 &  $\Lambda_2^{(0)\dagger} D T_2^{(1)}$ & 8.14 \\
$T_2^{(1)\dagger} D T_2^{(0)}$ & 8.70 &  $\Lambda_2^{(1)\dagger} D T_2^{(0)}$ & 7.07 \\
Others &  11.30       & Others & $-8.36$ \\
\hline \hline \\
\end{tabular}
\label{tab2}
\end{table}

Because of the non-Hermitian nature of $\Bar{H}$, it also exhibits a left eigenvector with the eigenvalue equation as
\begin{equation}
\label{eq:21}
\langle \Phi_{0}|\hat L_{k}\Bar{H} = \langle \Phi_{0}|\hat L_{k} E_{k} ,
\end{equation}
where $\hat{L}_{k}$ is a linear de-excitation operator with a form of
\begin{equation}
\label{eq:22}
\hat{L}_{k}= l_{0}+\sum_{i,a} l_{a}^{i}\;\hat{a}_{i}^{\dagger}\hat{a}_{a}+\sum_{i<j\atop a < b} l_{ab}^{ij}\;\hat{a}_i^{\dagger}\hat{a}_j^{\dagger} \hat{a}_b \hat{a}_{a}+\cdots .
\end{equation}
Like the LR approach, the two sets of eigenfunctions of the EOM-RCC method follow the biorthogonality condition, such as
\begin{equation}
\label{eq:23}
\langle \Phi_{0}|\hat L_{k}\hat{R}_{l}|\Phi_{0} \rangle = \delta_{kl} .
\end{equation}
Also, the EOM-RCCSD method often fails to provide quantitative accuracy, necessitating the inclusion of triples and higher excitations in the calculation. The full inclusion of triples in the EOM-RCC method (EOM-RCCSDT) scales as iterative $O(N^8)$ with a storage requirement that scales as $O(N^{6})$. Partial triples correction schemes are also defined for the EOM-RCC method (EOM-RCC3), where $\hat{T}$ and $\hat{R}$ operators are truncated using perturbative arguments. The EOM-RCC3 method scales as iterative $O(N^7)$ with a storage requirement similar to that of the EOM-RCCSD method. 

\begin{table}[t] 
\centering
\begin{threeparttable} 
\caption{Ground-state $\alpha$ value (in a.u.) of at different levels of calculations using molecular code. The values shown in bold fonts are considered in the analysis for final recommendation.}
\begin{tabular}{lcc}
\hline \hline \\
Code & Level of theory & $\alpha$ (a.u.)\\
\hline \\
Molecular & FF-DHF/d-aug-dyall-ae2z &  232.64\\
Molecular & FF-DHF/d-aug-dyall-ae3z  & 232.67\\
Molecular & FF-DHF/d-aug-dyall-ae4z &  \textbf{232.69}\\
Molecular & FF-RCCSD/s-aug-dyall-ae2z \tnote{a}  & 204.64 \\
Molecular & FF-RCC3/s-aug-dyall-ae2z\tnote{a} & 201.35\\
Molecular & FF-RCCSDT/s-aug-dyall-ae2z\tnote{a}  & \textbf{201.19} \\
Molecular & FF-RCCSDT/s-aug-dyall-ae2z\tnote{b}  & 202.15  \\
Molecular & FF-RCCSDT(Q)/s-aug-dyall-ae2z\tnote{b}  & 201.90  \\
Molecular & NR-FF-CCSD/d-aug-dyall-ae2z  & 196.39 \\
Molecular & FF-RCCSD/d-aug-dyall-ae2z\tnote{a}  & 203.26 \\
Molecular & FF-RCCSD/d-aug-dyall-ae3z\tnote{a} & 203.01 \\
Molecular & FF-RCCSD/d-aug-dyall-ae4z\tnote{a}  & \textbf{202.45} \\
Molecular & FF-RCC3/d-aug-dyall-ae2z\tnote{a}  &  200.42 \\
Molecular & LR-RCCSD/d-aug-dyall-ae2z\tnote{a}  &  196.39\\
         \hline \hline 
    \end{tabular}
        \label{table3}
    \begin{tablenotes}
    \small 
    \item [a]18 electrons frozen.
    \item [b]18 electrons frozen and a virtual orbital energy cutoff at  50 a.u. has been used. 
    \end{tablenotes} 
    \end{threeparttable} 
\end{table}

\begin{table*}[!t]
\caption{Demonstration of convergence of the $\alpha$ values of the ground and excited states of Sr with respect to increase in the size of molecular basis functions. CBS(Martin) offers the extrapolated value from the complete set. As can be seen, the ground state value convergences faster than the excited state results.}
\begin{tabular}{lccccccc}
\hline\hline
\text{Basis}
& \multicolumn{7}{c}{\text{Polarizability (a.u.)}} \\
\cline{2-8}
&
$^1S_0$
& $^3P_0$
& \multicolumn{2}{c}{$^3P_1$}
& \multicolumn{3}{c}{$^3P_2$} \\
\cline{4-5} \cline{6-8}
&
&
&
$|M_J|=0$ & $|M_J|=1$
& $|M_J|=0$ & $|M_J|=1$ & $|M_J|=2$ \\
\hline

d-aug-dyall.ae2z
& 203.26 & 344.79 & 312.99 & 371.98 & 410.84 & 389.07 & 326.12 \\

d-aug-dyall.ae3z
& 203.01 & 418.62 & 384.50 & 453.93 & 506.76 & 481.48 & 408.81 \\

d-aug-dyall.ae4z
& 202.45 & 434.84 & 400.20 & 472.37 & 529.07 & 502.97 & 427.96 \\

CBS (Martin)
& 201.67 & 443.36 & 408.44 & 482.48 & 541.93 & 515.33 & 438.92 \\
\hline\hline
\end{tabular}
\label{tab4}
\end{table*}

\section{Results and discussion}

\subsection{Overview of past results}

Before discussing the results of our calculations, it is instructive to first review the history of theoretical and experimental determinations of the static dipole polarizability of the Sr ground state, $5s^{2},{}^{1}S_{0}$. Table \ref{tab:full_sr_results_new} summarizes the reported ground- and excited-state polarizabilities of Sr over the years. As evident from the table, the experimental determination of the ground-state $\alpha$ value is associated with a relatively large uncertainty, while a wide range of theoretical values has been reported, generally lying within the experimental error bar. The experimental measurement was performed in 1974 by Schwartz, Miller, and Bederson~\cite{Schwartz1974}, who obtained $\alpha = 186 \pm 15$~a.u. using an atomic-beam electric-field balance (E--H balance). This relatively large experimental uncertainty has not been substantially reduced since then, leaving theoretical calculations as the primary source of accurate estimates of the Sr ground-state polarizability. Between 2004 and 2015, theoretical values in the range of approximately $187$--$202$~a.u. were reported using a variety of many-body approaches, including relativistic Douglas--Kroll CCSD(T)~\cite{LimSchwerdtfeger2004}, hybrid configuration-interaction plus many-body perturbation theory (CI$+$MBPT) sum over states approach~\cite{PhysRevA.74.020502}, all-order coupled-cluster calculations combined with experimental matrix elements \cite{MitroySafronovaClark2010}, configuration interaction with core polarization~\cite{MitroyZhang2010}, Dirac MBPT$+$CCSD \cite{SinghSahooDas2013}, perturbed relativistic coupled-cluster (PRCC) methods \cite{BKSahoo2,BKSahoo6,Chattopadhyay2014}, and large-basis CCSD(T)/CCSDT$+$CI calculations~\cite{Yu2015}. More recently, an augmented PRCC calculation incorporating QED corrections, including VP and self-energy contributions, together with perturbative triples and based on the Dirac--Coulomb--Breit Hamiltonian, reported a notably lower value of $\alpha = 176.50 \pm 2.65$~a.u.~\cite{PhysRevA.113.032817}.

To highlight the wide range of theoretical values reported over the last five decades, we plot them together with the available experimental result in Fig.~\ref{fig:sr-pol-history}. The figure shows that the theoretical values at the two extremes deviate by more than seven standard deviations from the long-standing recommended value of $197.2 \pm 0.2$~a.u. \cite{MitroySafronovaClark2010}, although most of the theoretical predictions remain within the broad uncertainty interval of the 1974 experimental measurement. The substantial spread among the theoretical results, with several calculations lying outside the quoted uncertainty intervals of one another, underscores the need for a systematic multi-method investigation accompanied by a rigorous assessment of the associated uncertainties. The accuracy of an {\it ab initio} calculation of an atomic property using many-body methods is primarily governed by three key factors: (i) the ability of the employed many-body method to capture electron-correlation effects, (ii) the completeness of the one-electron basis set, and (iii) the level of approximation adopted in the Hamiltonian. Ideally, convergence should be established with respect to all three aspects before a reliable uncertainty can be assigned to the calculated property. From the atomic Hamiltonian perspective, incorporating as many relevant relativistic and quantum-electrodynamic (QED) effects as practicable is desirable for achieving high accuracy. In practice, however, the DC Hamiltonian provides a suitable starting point for Sr, with the Breit and QED interactions included as corrections for high-precision polarizability calculations, where necessary~\cite{BKSahoo5,BKSahoo8}.

For the treatment of electron correlation, the LR and FF RCC approaches employed in the present work provide complementary and rigorous frameworks for evaluating the ground-state $\alpha$ of Sr. For the excited states, the EOM-RCC method is better suitable than the previously employed methods. Although computationally more demanding~\cite{10.1063/1.3518712,10.1063/5.0289155,10.1021/acs.jctc.5c01791}, EOM-RCC provides a flexible and systematically improvable framework compared with Fock-space RCC or CI$+$MBPT approaches. Due to use of the reduced matrix elements, we are able to  consider excitations of all electrons using atomic codes. However, we perform calculations using molecular codes by systematically including enlarged sets of one-electron basis functions to examine the convergence of the $\alpha$ values with respect to basis-set completeness~\cite{10.1063/1.3518712}. Together, these analyses allow us to assess the major sources of uncertainty associated with electron correlation, basis-set completeness, and the Hamiltonian, thereby providing a more reliable estimate of the Sr polarizabilities. We first analyze the ground-state polarizability of Sr using the RCC methods discussed in the previous section, employing both the atomic and molecular codes. The results for the excited states are presented subsequently.

\subsection{Results from atomic codes}

Table~\ref{tab1} summarizes the calculated ground-state polarizability $\alpha$ of Sr at different levels of approximation using the DC Hamiltonian, together with higher-order relativistic corrections evaluated using the RCCSD method with the atomic code. At the lowest-order DHF level, we obtain $\alpha=156.91$~a.u.. Inclusion of orbital-relaxation effects through the RPA (equivalently referred to as core-polarization effects to all-orders) increases the polarizability substantially to $232.55$~a.u.. This large change demonstrates the important role of orbital relaxation (core polarization) effects in determining the polarizability of Sr. The RCCSD method, which incorporates core-polarization and pair-correlation effects, together with their mutual correlation, to all-orders at the singles-and-doubles excitation level, yields $\alpha=197.24$~a.u.. Inclusion of perturbative triple excitations through the RCCSD(T) approximation reduces this value to $195.25$~a.u.. In contrast, the RNCCSD method gives a substantially lower value of $185.55$~a.u.. To understand the origin of this difference between the RCCSD and RNCCSD results, we analyze the contributions from the individual terms in the two formalisms in Table~\ref{tab2}.

As evident from Table~\ref{tab2}, the dominant contributions in both approaches arise from the $DT_1^{(1)}$ term and its counterpart. In the RCC formulation, these two terms contribute equally, $114.41$~a.u. each, owing to the symmetric treatment of the ket and bra states in the expectation-value expression. In the RNCC formulation, however, the corresponding left-hand contribution, $\Lambda_1^{(1)\dagger}D$, is only $83.77$~a.u., substantially smaller than the contribution from the $T_1^{(1)\dagger}D$ term. In addition, the cross-correlation terms involving $T_1^{(0/1)}$ and $T_2^{(0/1)}$ operators together with the dipole operator make significant negative contributions. In particular, $T_2^{(0)\dagger}DT_1^{(1)}$ and its counterpart contribute $-14.84$~a.u. each in the RCC formulation. These terms represent the interplay between the ground-state electron correlation and the first-order response to the external electric field and therefore play an important role in determining the final value of $\alpha$. The category denoted by ``Others'', which includes the remaining higher-order correlation contributions, gives $+11.30$~a.u. in the RCC formulation, whereas it contributes $-8.36$~a.u. in the RNCC formulation. This reversal in sign, together with the differences in the dominant terms, accounts for a substantial part of the discrepancy between the RCCSD and RNCCSD results. The comparison demonstrates that the calculated ground-state polarizability is sensitive to the detailed treatment of correlation and, in particular, to the formulation used for evaluating the expectation value.

It is also worth noting that the RNCCSD expression contains fewer explicit terms than the RCCSD expectation-value expression. Some of the nonlinear RCCSD contributions effectively incorporate effects associated with higher level excitations. The higher-order excitation contributions are expected to converge more rapidly in the RNCC framework than in the conventional RCC expectation-value approach. Inclusion of additional higher-order excitations, particularly triple excitations, would therefore be desirable for further improving the RNCCSD result. Such calculations, however, are beyond the present capabilities of our atomic code. We therefore estimate the residual uncertainty by considering the available higher-order corrections. Nevertheless, the Breit and VP contributions to the ground-state polarizability are found to be $-0.06$ and $-0.01$~a.u., respectively, which are negligibly small. Therefore, accuracy of the $\alpha$ value mostly rely on the correct treatment of electron correlation effects. Together with the estimated contribution from partial triple excitations and the remaining correlation and basis-set uncertainties, we conservatively assign an overall uncertainty of approximately $5.0$~a.u. to the ground-state polarizability. Taking the RCCSD(T) value as the central estimate, we recommend the $\alpha$ value of the ground state of Sr to be $195(5)$  a.u..

\subsection{Result summary from molecular codes}

In Table~\ref{table3}, we summarize the calculated ground-state polarizability obtained using the molecular codes at different levels of theory and basis sets. All calculations were performed with 18 frozen electrons and the full virtual space, except for the RCCSDT(Q) calculation, which was performed with 28 frozen electrons and a virtual cutoff of 50 a.u.. The first thing to be noticed from the FF approach of the molecular code is, the DHF values from all the three considered basis sets are consistent and they agree quite well with the RPA result from the atomic code, $232.55$~a.u., as shown in Tables~\ref{tab1}. This provides an important consistency check between the FF and LR implementations. Large difference seen between the RCCSD and RNCSSD values of the ground state $\alpha$ from atomic code illustrates the sensitivity of accurate treatment of correlation and wave-function normalization. It is, therefore, necessary to closely follow the electron correlation propagation from singles and doubles approximations to triples effects in the calculations. Unlike the LR approach, contributions from individual RCC terms cannot be obtained in the FF approach. However, the FF results can be more reliable due to its inherent behavior of including orbital relaxation effects, provided the numerical differentiation adopted in this approach is stable.   

As can be seen from the above table, the FF-RCCSD results show only mild sensitivity to the basis set. The ground state $\alpha$ value decreases from 204.64 a.u. to 203.26 a.u. when going from the s-aug-dyall-ae2z to the d-aug-dyall-ae2z basis set. On the other hand, the value further decreases from 203.26 a.u. to $202.45$ a.u. when going from d-aug-dyall-ae2z to d-aug-dyall-ae4z. The inclusion of perturbative triples via FF-RCC3 lowers the value to 201.35 a.u. at the s-aug-dyall-ae2z level. The full iterative FF-RCCSDT calculation yields $201.19$ a.u. at the same basis set. The FF-RCCSDT(Q) value shows a further decrease of 0.25 a.u.. A notable observation is the difference of 6.87 a.u. between the FF-RCCSD and LR-RCCSD values in the d-aug-dyall-ae2z basis. This is due to the neglected orbital relaxation effects in the LR-RCCSD  formulation \cite{10.1063/5.0274450}. This is confirmed by performing a non-relaxed FF (NR-FF-RCCSD) calculation by including the field after the DHF step, which gives identical results to those of the LR-RCCSD method. The LR-RCCSD value of $196.39$ a.u. at the d-aug-dyall-ae2z basis is in close agreement with the atomic RCCSD result of $197.24$ a.u. in the same approach. This confirms the mutual consistency of the FF and LR RCCSD approaches, and provides consistency check in the reliability of our molecular code results for the excited states.

The comparison of our composite polarizability values with the previous theoretical and experimental data is presented in Table~\ref{tab:full_sr_results}. Our ground-state polarizability of $198.24$ a.u. is in reasonable agreement with the values reported by other groups, including 199.7 a.u.\ and 197.2 a.u.\ from different RCC-based approaches. The excited-state polarizability  corresponding to the clock transition $446.86$ a.u. $^{3}$P$_{0}$ shows a larger spread among different theoretical predictions, ranging from $410$~a.u.~\cite{Guo_2010} to $497$~a.u.~\cite{Mitroy10082010} in the literature~\cite{PhysRevA.74.020502,PhysRevA.78.032508,Guo_2010,Mitroy10082010,PhysRevA.87.012509,PhysRevA.92.052515}, which underscores the difficulty of obtaining converged excited-state polarizabilities for heavy atoms and the sensitivity of this quantity to the computational approach. Our excited-state polarizability value for the $^{3}$P$_{1}$ state is 487 a.u. which lies between the two previously reported values~\cite{PhysRevA.92.052515,PhysRevA.78.032508} of 480.9 a.u. and 498.8 a.u. To the best of our knowledge, only one previously reported value~\cite{PhysRevA.92.052515} is available for the $^{3}$P$_{2}$ state, and our recommended value of 548 a.u. is in good agreement with that.

\begin{table}[t]
\centering
\caption{Demonstration of the effects of the higher-order correlations and relativistic effects to the CBS results of the of the $^{1}$S$_{0}$ and $^{3}$P$_{0}$ state polarizabilities of the Sr atom. The roles of the inner-core electron correlations are also shown explicitly.}
\begin{tabular}{ c c c}
\hline \hline \\
  & \multicolumn{2}{c}{Polarizability (in a.u.)}   \\
 \cline{2-3}  \\ Method &$^{1}$S$_{0}$     & $^{3}$P$_{0}$  \\
 \hline \\
 RCCSD   &   201.67  &  443.36    \\
 (CBS--DZ/TZ/QZ) &&\\
  \\
 + $\Delta{T}$ correction & $-3.45$  & +0.65    \\
  (s-aug-dyall.ae2z)    &        &                  \\
  \\
   $\Delta{(T_{RCC3})}$ correction & $-3.38$  & +0.26    \\
  (s-aug-dyall.ae2z)    &        &                  \\
  \\
    $\Delta{(T_{RCC3})}$ correction & $-4.01$  & +1.12    \\
  (d-aug-dyall.ae2z)    &        &                  \\
  \\
 + $\Delta$core correction & $-0.02$   & +1.42      \\
  (d-aug-dyall.ae2z) & &  \\
  \\
 + $\Delta$Gaunt correction & +0.04    & +1.43       \\
  (d-aug-dyall.ae2z)    &        &                  \\
  \\
   $\Delta$Breit correction & +0.02    & +1.25       \\
  (d-aug-dyall.ae2z)    &        &                  \\
  \\
 Composite Value$^{\dagger}$  &   198.24   &   446.86    \\ 
\hline \hline 
\end{tabular}
\\
$^{\dagger}$Composite = RCCSD(CBS) + $\Delta T$ + $\Delta$core + $\Delta$Gaunt. The rest of the correction is shown for comparison only and is not included in the composite.
\label{table:3}
\end{table}

\subsubsection{Effect of basis functions}

Table~\ref{tab4} presents the polarizability values for the ground $^1S_0$ state and the excited triplet $^3P_0$, $^3P_1$ and $^3P_2$ states of the Sr atom using different basis sets. The polarizability values were calculated using the finite difference of the total energy obtained in the presence and absence of an electric field. For testing purposes, four different field strengths ranging from 0.01 to 0.0001 atomic units were selected. The change from 0.001 a.u. to 0.0001 a.u. is found to be very small. The remaining polarizability calculations in this article are performed at a field strength of 0.001 a.u. Inclusion of a larger number of diffuse functions helps improve the accuracy of calculated $\alpha$ values in the many-body methods, and the doubly augmented dyall-basis has been reported to give the best compromise between computational cost and accuracy \cite{10.1063/5.0274450}. We have performed the calculations using an uncontracted d-aug-dyall.aeXZ ($X = 2, 3, \text{and } 4$) basis set with a frozen core approximation (keeping eighteen electrons frozen for the calculation) to investigate if the deviation in results could be attributed to the utilization of a finite-size basis set.
\begin{table*}[!t]
\caption{Comparison of CBS values (in a.u.) of the ground excited state polarizabilities of Sr using Martin, Feller+Helgeker, Feller+Lesiuk and Peterson+Dunning extrapolation schemes. See the text for a detailed discussion.}
\begin{tabular}{lccccccc}
\hline\hline
\text{Basis}
& \multicolumn{7}{c}{\text{Polarizability (a.u.)}} \\
\cline{2-8}
&
$^1S_0$
& $^3P_0$
& \multicolumn{2}{c}{$^3P_1$}
& \multicolumn{3}{c}{$^3P_2$} \\
\cline{4-5} \cline{6-8}
&
&
&
$|M_J|=0$ & $|M_J|=1$
& $|M_J|=0$ & $|M_J|=1$ & $|M_J|=2$ \\
\hline

EOM-RCCSD (2z/3z/4z)-Martin 
& 201.67 & 443.36 & 408.44 & 482.46 & 541.93 & 515.33 & 438.92 \\

EOM-RCCSD (2z/3z/4z)-Feller + Helgaker
& 201.83 & 446.01 & 411.01 & 485.26 & 544.95 & 518.25 & 441.55 \\

EOM-RCCSD (2z/3z/4z)-Feller + Lesiuk
& 201.60 & 445.11 & 410.14 & 484.46 & 544.34 & 517.65 & 440.98 \\

EOM-RCCSD (2z/3z/4z)-Peterson + Dunning
& 202.07 & 443.25 & 408.34 & 482.00 & 540.83 & 514.28 & 438.03 \\
\hline\hline
\end{tabular}
\label{tab:cbs_methods}
\end{table*}

The extrapolation to the complete basis set (CBS) limit is performed using the Martin's extrapolation formula \cite{MARTIN1996669}. The CBS extrapolation can be major source of potential errors in our calculations and will be discussed in detail in the error estimation section below. The polarizability of the ground state shows a slow but steady decrease with the increase in basis set size, changing only by $\sim 1.6$ a.u. across the entire range from double-zeta to the CBS limit.

On the other hand, the excited $^3P_0$ state polarizability exhibits a remarkably different behavior. The value increases dramatically from 344.79 a.u. at d-aug-dyall.ae2z to 434.84 a.u. at d-aug-dyall.ae4z, and further to 443.36 a.u. at the CBS limit. The change from the double-zeta to the triple-zeta basis set is 73.83 a.u., and the subsequent change upon going to d-aug-dyall.ae4z is a smaller 16.22 a.u. This much slower convergence of the excited-state polarizability compared to the ground state can be attributed to the more diffuse nature of the excited-state wave function, which demands larger and more flexible basis sets for an adequate description. Nevertheless, even with the d-aug-dyall.ae4z basis set, the excited-state polarizability did not converge, and CBS extrapolation leads to a value of 443.36 a.u. Therefore, the CBS extrapolation at the EOM-CCSD level within the RCC theory framework gives the ground-state $\alpha$ value that differs from the experimental value by $\sim 16$ a.u.. The remaining discrepancy is presumably due to correlation effects from higher-level excitations and omitted relativistic corrections; these must be accurately estimated to obtain a precise ground-state $\alpha$, especially given that the experimental value itself carries an error bar of 15 a.u.. There are no experimental values of $\alpha$ for the excited state  available for comparative analysis.

Table~\ref{tab4} also shows detailed basis set convergence behavior for other excited states, including the $^3P_1$ and $^3P_2$ states with their $|M_J|$ sublevels. The $^3P_1$ and $^3P_2$ states show similar trends to $^3P_0$: their $\alpha$ values increase substantially with the basis set size, and the $|M_J|$-dependent values show significant spread, reflecting the anisotropic nature of the excited-state charge distributions. For example, the $^3P_2$ state CBS $\alpha$ values range from 438.92 ($|M_J|=2$) to 541.93 ($|M_J|=0$) a.u., demonstrating the strong $M_J$-dependence of the excited-state $\alpha$ values.

\subsubsection{Roles of electron correlation effects}

To observe the impact of the higher correlation, we integrate the triples excitations by conducting calculations using the RCCSDT method. The calculation of the unaccounted correlation contributions to the $\alpha$ values is presented in the following way:
\begin{equation}
\label{eq:24}
\Delta{T}_{(\alpha)} = \text{RCCSDT}_{(\alpha)} - \text{RCCSD}_{(\alpha)} ,
\end{equation}
where the subscripts $\alpha$ stand for the polarizability at a particular basis set arising from the triple excitations using the RCCSDT method. The uncontracted s-aug-dyall.ae2z basis set was employed for evaluating the $\Delta T$ correction with the frozen-core approximation.

Table~\ref{table:3} clearly demonstrates that the inclusion of full triples has a notable impact on the polarizabilities. The inclusion of triple corrections leads to a decrease of 3.45 a.u. in the ground-state polarizability, while the excited-state polarizability increases by 0.65 a.u.. We have also computed the perturbative triples correction, $\Delta(T_{RCC3})$, which approximates the effect of connected triples through a perturbative correction. The $\Delta(T_{RCC3})$ correction yields $-3.38$ a.u. for the ground-state polarizability and $0.26$ a.u. for the excited-state polarizability, which closely resemble the values of the full RCCSDT. It should be noted that the effect of the approximated triples inclusion is slightly larger when one uses the d-aug-dyall.ae2z basis set, indicating that the effect of higher-order correlation is not strictly additive. In the final composite values, we adopt the $\Delta T$ correction from the full RCCSDT calculation, as it is more reliable than the perturbative estimate, and the non-additivity of the higher-order correlation is included in the uncertainty estimate. 


\subsubsection{Inner-core correlation and higher-order relativistic effects}

To reduce the computational cost, we employ the frozen-core approximation throughout, accepting a slight compromise in accuracy. The d-aug-dyall.aexz family of basis sets are all-electron basis sets; however, when used with the frozen-core approximation, the inner-core correlations are neglected. Consequently, to assess the specific impact of correlation effects attributed to the frozen inner eighteen electrons in our analysis, we carried out two sets of calculations. First, we considered all 38 electrons of the Sr atom using an uncontracted d-aug-dyall.ae2z basis set, designed to take care of the core correlation. Second, we repeated the calculation with the d-aug-dyall.ae2z basis while freezing the inner eighteen electrons. The difference in the results was considered as $\Delta$core and was added to our CBS results. It is evident from Table~\ref{table:3} that the impact of the core electron correlations on the ground-state polarizability is negligible ($-0.02$ a.u.). However, for the $^3P_0$ excited-state polarizability, the core correction is more significant at $+1.42$ a.u..

It is also important to investigate the influence of higher-order relativistic corrections on the polarizability. To account for the Gaunt contribution, the Hamiltonian is augmented by incorporating the Gaunt operator [as defined in Eq.~(\ref{eq:11})]. It should be noted that the Gaunt correction is added only at the level of DHF wave-function evaluation, but it is not included in the inclusion of correlation effects. Subsequently, a EOM-RCCSD calculation is executed using an uncontracted d-aug-dyall.ae2z basis set. Repeating the same procedure without the Gaunt operator in the Hamiltonian and subtracting the results of both calculations yields the $\Delta$Gaunt correction. Based on the data presented in Table~\ref{table:3}, the ground-state polarizability changes by only +0.04 a.u., while the excited-state polarizability shows a more noticeable change of +1.43 a.u. However, the inclusion of the full Breit correction does not lead to any appreciable change in the results compared to the Gaunt-only correction. The deviation caused by the gauge (retardation) term in the $^3P_0$ excited-state polarizability is only $-0.18$ a.u.. 

Upon incorporating all the necessary corrections into the CBS result, the composite values for the ground- and excited-state polarizabilities are 198.24 a.u. and 446.86 a.u., respectively. The ground-state polarizability from the EOM-RCC composite value (198.24 a.u.) is in reasonable agreement with the atomic RCCSD result (197.24 a.u.) and lies within the uncertainty of the recommended atomic value of 195(5) a.u.. This indicates that our {\it ab initio} approach achieves high-accuracy results. One can expect similar accuracy for the other excited states reported in Table~\ref{tab:full_sr_results} once convergence is achieved with respect to the one-electron basis set, level of correlation, relativistic effects, and core correlation in the EOM-RCC calculations.

\begin{table}[!t]
\centering

\caption{Recommended $\alpha$ values (in a.u.) to the investigated ground excited states of Sr. For the recommendation, results at the RCCSD method approximation are considered as the central values and other corrections such as from triples ($\Delta$T), inner core ($\Delta$Core), and Gaunt term ($\Delta$Gaunt) are taken into account. }
\label{tab:full_sr_results}

\begin{tabular}{@{}l r r r r@{}}
\hline\hline

\text{Source}& \multicolumn{4}{c}{\text{$\alpha$ value (a.u.)}}  \\
\cline{2-5} \\
 & \multicolumn{1}{c}{$^1S_0$} & \multicolumn{1}{c}{$^3P_0$} &
\multicolumn{1}{c}{$^3P_1$} & \multicolumn{1}{c}{$^3P_2$ } \\
              &  &  & ($|M_J|=1$) &  ($|M_J|=0$) \\
\hline \\
\text{Main (RCCSD)} & 201.67 & 443.36 & 482.48 & 541.93   \\
\text{$\Delta$T} & $-3.45$ & $+0.65$ & $+1.54$ & $+2.6$   \\
\text{$\Delta$Core} & $-0.02$ & $+1.42$ & +1.68& $+2.14$  \\
\text{$\Delta$Gaunt} & $+0.04$ & $+1.43$ & $+1.54$ & $+1.5$  \\
\hline \\
Recommended & \text{198(1)} & \text{447(4)} & \text{487(5)} & \text{548(6)} \\
\hline \hline
\end{tabular}
\end{table}

\subsubsection{Error estimation}

The contribution of all different corrections added to the polarizabilities is analyzed to estimate the uncertainties in the final results. It is clear that the major contribution to the uncertainty comes from the neglected triples corrections and the basis set incompleteness for the excited-state $\alpha$ values. The first source of uncertainty due to triples is because of the EOM-RCCSDT values that are available only at the s-aug-dyall.ae2z basis set, and therefore the basis set dependence of the triples correction cannot be directly assessed. However, the differences between the RCC3 and RCCSD results across different basis sets provide estimates of the variance of the RCCSDT corrections with respect to the basis set. The differences between the $\Delta T$ and $\Delta(T_{RCC3})$ corrections with the s-aug-dyall.ae2z and d-aug-dyall.ae2z basis sets are 0.63 a.u. and 0.86 a.u. for the ground and excited-state $\alpha$ values, respectively. We take these differences as the uncertainty due to non-additivity of the triples correction across the basis sets.

\begin{table*}[t]
\centering
\renewcommand{\arraystretch}{1.15}
\setlength{\tabcolsep}{6pt}
\caption{Estimated net and fractional BBR shift values for the clock and other transitions involving the ground and excited states in Sr.}
\label{tab:bbr_shift}

\begin{tabular}{@{}l l r r r@{}}

\hline \hline
\text{Sr transition} &
\text{Source} &
\multicolumn{1}{c}{$\delta\nu$ (Hz)} &
\multicolumn{1}{c}{$\nu_0$ (Hz)} &
\multicolumn{1}{c}{$\delta\nu / \nu_0$} \\
\hline\\

\multirow{2}{*}{$^1S_0 \rightarrow {}^3P_0$}
& This work
& $-2.322$ & $4.29\times 10^{14}$ & $-5.40\times 10^{-15}$ \\

& Previous work
& $-1.96(37)$ \cite{Guo_2010} &&$-4.56\times 10^{-15}$\\

& & $-2.09$ \cite{PhysRevA.92.052515} &&$-4.86\times 10^{-15}$\\
&   & $-2.354(32)$ \cite{PhysRevA.74.020502} &&$-5.5\times 10^{-15}$\\
 & & & \\ 
 $^1S_0 \rightarrow {}^3P_1$ ($|M_J|=1$) & This work
& $-2.838$ & $4.33\times 10^{14}$ & $-6.52\times 10^{-15}$ \\
 & & & \\
$^1S_0 \rightarrow {}^3P_2$ ($|M_J|=0$)
& This work
& $-3.241$ & $4.45\times 10^{14}$ & $-7.25\times 10^{-15}$ \\
\hline
\end{tabular}
\end{table*}

Contributions from higher-order excitations (quadruples and beyond) can be estimated from the difference between RCCSDT and RCC3 polarizability values. Twice this difference provides a conservative estimate of the quadruple and higher-level excitation correction, found to be 0.14 a.u. and 0.78 a.u. for the ground and first excited states, respectively. The magnitude of the perturbative quadruples (Q) correction for the ground state is 0.25 a.u., which confirms the reliability of our uncertainty estimate for the missing higher-order correlations. The next contributor to the uncertainty in the polarizability is the incompleteness of the one-electron basis set. It is actually the biggest contributor to the uncertainty for the excited-state polarizability. As one can see, the excited-state results have not converged even at the quadruple-zeta basis set. So, one needs to extrapolate it to the CBS limit. There is no unique way to extrapolate it to the CBS limit. We have tested four most popular approaches for basis set extrapolation, which are described below.

In all schemes, the extrapolation is carried out on the energies rather than directly on the polarizabilities. At each applied field strength, the total energy is partitioned as $E_X = E^{\rm DHF}_X + E^{\rm corr}_X$, where $X=2$, $3$, and $4$ denotes the cardinal number of the d-aug-dyall.ae$X$z basis set. The DHF and correlation energies are extrapolated separately to the CBS limit, and the extrapolated total energy, $E_{\rm CBS} = E^{\rm DHF}_{\rm CBS} + E^{\rm corr}_{\rm CBS}$, is then used in the finite-field expression of Eq.~(\ref{eq:3}) to obtain the CBS polarizability.

In the scheme of Martin~\cite{MARTIN1996669}, both the DHF and correlation energies are extrapolated using the three-parameter inverse-power expansion
\begin{equation}
E_X = E_{\rm CBS} + \frac{A}{(X+1/2)^4} + \frac{B}{(X+1/2)^6},
\label{eq:cbs_martin}
\end{equation}
which is solved exactly with the 2z, 3z, and 4z energies. In the  Feller+Helgaker and  Feller+Lesiuk schemes, the DHF energy is extrapolated with the three-point exponential form proposed by Feller~\cite{10.1063/1.462652},
\begin{equation}
E^{\rm DHF}_X = E^{\rm DHF}_{\rm CBS} + B\, e^{-CX}.
\label{eq:cbs_feller}
\end{equation}
In the Feller+Helgaker scheme, the correlation energy is extrapolated with the two-point $X^{-3}$ formula of Helgaker {\it et al.}~\cite{10.1063/1.473863},
\begin{equation}
E^{\rm corr}_{\rm CBS} = \frac{X^3E^{\rm corr}_X-(X-1)^3E^{\rm corr}_{X-1}}{X^3-(X-1)^3},
\label{eq:cbs_helgaker}
\end{equation}
while in the Feller+Lesiuk scheme it is extrapolated with the Riemann zeta-function based formula of Lesiuk and Jeziorski~\cite{doi:10.1021/acs.jctc.9b00705},
\begin{equation}
E^{\rm corr}_{\rm CBS} = E^{\rm corr}_X + X^4\left[\zeta(4)-\sum_{l=1}^{X} l^{-4}\right]\left(E^{\rm corr}_X-E^{\rm corr}_{X-1}\right),
\label{eq:cbs_lesiuk}
\end{equation}
with $\zeta(4)=\pi^4/90$. In both two-point formulas, the 3z and 4z energies are employed ($X=4$).  In the Peterson+Dunning scheme, the DHF energy is extrapolated with the mixed exponential/Gaussian form of Peterson, Woon, and Dunning~\cite{10.1063/1.466884},
\begin{equation}
E_X = E_{\rm CBS} + B\, e^{-(X-1)} + C\, e^{-(X-1)^2},
\label{eq:cbs_dunning}
\end{equation}
and the correlation energy is extrapolated with the same form, using the 2z, 3z, and 4z energies.

Table \ref{tab:cbs_methods} presents the $\alpha$ values in various basis set extrapolation schemes. The ground-state $\alpha$ shows a small spread of 0.47 a.u. across the four CBS schemes. The excited-state $\alpha$ values, on the other hand, give a much larger spread. A spread of  $^3P_0$ of 2.76 a.u., 2.46 a.u. and 3.27 a.u. has been observed for the $^3P_0$,$^3P_1$  and $^3P_2$  state, respectively. It is interesting to note that in all the cases, the Feller-Helgaker scheme gives the largest excited state polarizability, and the Peterson-Dunning scheme gives the smallest polarizability values. The mixed exponential-Gaussian Peterson formula assumes a rapid convergence behavior where the property corrections vanish exponentially as the cardinal number (L) increases. As a result, when applied to monotonically increasing properties like dipole polarizability, the Peterson formula predicts that finite basis set values are already near their convergence ceiling, thereby yielding a systematically smaller and more conservative complete basis set limit compared to the slow-decaying $L^{-3}$dependent Helgaker model. Half of the spread of polarizability values has been considered to be the uncertainty due to finite basis set. 

The maximum magnitude of the uncertainty for the frozen-core approximation and higher-order relativistic effects can be taken to be half of their total contributions.  The difference between the Gaunt and Breit interactions is taken as the uncertainty due to higher-order relativistic effects. Another source of error can be numerical noise. We find the optimal value of the field strength is 0.001 a.u..  We added them to the error estimate as errors due to numerical uncertainty. 

In Table~\ref{tab:full_sr_results}, we report the uncertainties in the calculated $\alpha$ values for both the ground and excited states in parentheses alongside the corresponding final results.  As can be seen, the ground-state polarizability of $198.24$ a.u. lies within the error bar of the experimental value $186 \pm 15$ a.u. \cite{PhysRevA.54.1973}, given in Table \ref{tab:full_sr_results_new}. Our recommended value for ground-state polarizability based on the EOM-CC composite is $198.24 \pm 1.14$ a.u., while the atomic RCC calculations recommend a value of $195 \pm 5$ a.u. Both values are consistent within their respective uncertainties and lie within the experimental error bar. The excited-state polarizability for the $^3P_0$,$^3P_1$  and $^3P_2$  states are $446.86 \pm 4.01$ a.u., $487.24 \pm 3.99$ a.u. and $548.17 \pm 4.62$ a.u., respectively. Although these results are in close agreement with the previous calculations listed in Table \ref{tab:full_sr_results_new}, they offer improved precision.

\subsection{BBR shift estimation}

The BBR shift for a transition between the states $i$ and $j$ can be estimated using the approximation \cite{PhysRevA.74.020502}
\begin{equation}
\label{eq:1}
\delta E= -\dfrac{1}{2}\left( 831.9\: V/m \right)^{2}\left( \dfrac{T}{300}\right)^{4}\left( \alpha^{0}_{i}-\alpha^{0}_{j}\right)\left( 1+\eta\right) ,
\end{equation}
where $\alpha^{0}_{i}$ and $\alpha^{0}_{j}$ are the static scalar polarizabilities of the $i$ and $j$ states, respectively, and $\eta$ is a small dynamic correction resulting from the frequency distribution of the BBR field. The $\alpha$ values presented in this work for the $^1S_0$ and $3P_0$ states correspond to the scalar static values owing to their zero total angular momentum.  

By neglecting the earlier found small $\eta$ value, $7 \times 10^{-4}$ \cite{PhysRevA.74.020502}, the BBR shifts to the $^1S_0 \rightarrow {}^3P_0$ clock transition and other transitions are presented in Table~\ref{tab:bbr_shift} and compared with the literature data. The table clearly indicates that our estimated BBR shift to the clock transition $^1S_0 \rightarrow {}^3P_0$ is $-2.322(50)$ Hz, which is consistent with but slightly smaller in magnitude than the previously calculated value of $-2.354(32)$ Hz by Lisdat {\it et al.}\ \cite{PhysRevResearch.3.L042036}. Also, the estimated relative BBR shift $\delta\nu/\nu_0 = -5.40 \times 10^{-15}$, is in good agreement with the earlier estimations. The $^1S_0 \rightarrow {}^3P_1$ ($|M_J|=1$) transition shows a BBR shift of $-2.838$ Hz with $\delta\nu/\nu_0 = -6.52 \times 10^{-15}$, and the $^1S_0 \rightarrow {}^3P_2$ ($|M_J|=0$) transition yields a BBR shift of $-3.241$ Hz with $\delta\nu/\nu_0 = -7.25 \times 10^{-15}$. The larger BBR shifts for these transitions reflect the larger differential polarizabilities. These values provide useful reference data for experiments involving the $^3P_1$ and $^3P_2$ states of Sr.

\section{Conclusion}

We have calculated very precise electric dipole polarizability values for the ground and first excited states of the Sr atom using relativistic coupled-cluster theory. Multiple approaches within the four-component relativistic framework, including the linear-response RCC, finite-field EOM-RCC, and perturbative RCC methods using both atomic and molecular codes, were employed to cross-validate the ground-state polarizability. The consistency among these independent calculations provides high confidence in the reliability of our results. The roles of finite basis size, electron correlation effects, and higher-order relativistic corrections to the electric dipole polarizability values are analyzed systematically. Our study shows that the incompleteness of the one-electron basis is a critical contributor to the error in the calculation, particularly for the excited-state polarizability, which converges very slowly with respect to the basis set size. The accuracy of the complete basis set result depends upon the choice of the basis set extrapolation scheme. The triples correction has a significant effect on the ground-state polarizability, reducing it by about 3.45 a.u. from the singles and doubles approximated RCC value. The core-correlation and Gaunt corrections are relatively small for the ground state but more significant for the excited-state polarizability. Our ground-state polarizability value of $198.24 \pm 1.14$ a.u. from the RCC composite lies well within the experimental error bar of $186 \pm 15$ a.u. This demonstrates the potential of our employed method to estimate the properties of atomic systems, the Sr atom in particular, precisely. Based on these analyses, we provide the recommended value for the excited-state dipole polarizability to be $446.86 \pm 4.01$ a.u.. The atomic RCC calculations additionally recommend a ground-state polarizability value of $195 \pm 5$ a.u., which is consistent with the molecular RCC result.

Using these polarizability values, we estimated the BBR shift of the $^1S_0 \rightarrow {}^3P_0$ clock transition to be $-2.322 \pm 0.05$ Hz, corresponding to a fractional frequency shift of $(-5.40 \pm 0.12) \times 10^{-15}$. This value is in good agreement with the earlier calculated value of $-2.354(32)$ Hz by Lisdat {\it et al}.. We have also provided BBR shift estimates for the $^1S_0 \rightarrow {}^3P_1$ and $^1S_0 \rightarrow {}^3P_2$ transitions, which serve as useful reference data for future experiments.

\section*{Acknowledgment}

The authors acknowledge the support from the ANRF under grant no. CRG/2023/002558, ISRO under its RESPOND program, Matrix project of DST-SERB, CSIR-India, DST-Inspire Faculty Fellowship, Prime Minister's Research Fellowship, IIT Bombay supercomputing facility, and C-DAC Supercomputing resources (PARAM Yuva-II, Param Bramha) for computational time. AKD acknowledges the research fellowship funded by the EU NextGenerationEU through the Recovery and Resilience Plan for Slovakia under project No 09I03-03-V04-00117. BKS also acknowledges the support from the Department of Space, Government of India and for using ParamVikram-1000 to carry out atomic calculation.

\section*{Conflict of interest}

The authors declare no competing financial interests.

\bibliography{main}
\end{document}